\documentclass[preprint]{aastex}
\usepackage{color}

\slugcomment{Not to appear in Nonlearned J., 45.}
\shorttitle{Effects of high-energy ionizing particles on the Si:As mid-infrared detector array on board the AKARI satellite}
\shortauthors{Mouri et al.}

\begin{document}
\title{Effects of high-energy ionizing particles on the Si:As mid-infrared detector array on board the AKARI satellite}
%\title{ionizing radiation effects on the Si:As mid-infrared detector array on board the AKARI satellite}
%\title{「あかり」搭載 Si:As 検出器に対する宇宙放射線の影響と補正方法の確立}
\author{
A. Mouri\altaffilmark{1},
H. Kaneda\altaffilmark{1},
D. Ishihara\altaffilmark{1},
S. Oyabu\altaffilmark{1},
M. Yamagishi\altaffilmark{1},
T. Mori\altaffilmark{1},
T. Onaka\altaffilmark{2},
T. Wada\altaffilmark{3}, and 
%H. Matsuhara\altaffilmark{3},
H. Kataza\altaffilmark{3}
}

\altaffiltext{1}{Graduate School of Science, Nagoya University, Furo-cho, Chikusa-ku, Nagoya, Aichi 464-8602, Japan}
\altaffiltext{2}{Department of Astronomy, Graduate School of Science, University of Tokyo, 7-3-1 Hongo, Bunkyo-ku, Tokyo 113-0033, Japan}
\altaffiltext{3}{Institute of Space and Astronautical Science (ISAS), Japan Aerospace Exploration Agency (JAXA), 3-1-1 Yoshinodai, Sagamihara,  Kanagawa 252-5210, Japan}

\begin{abstract}
We evaluate the effects of high-energy ionizing particles 
on the Si:As impurity band conduction (IBC) mid-infrared detector on board {\it AKARI}, 
the Japanese infrared astronomical satellite.
IBC-type detectors are known to be little influenced by ionizing radiation.
However we find that the detector is significantly affected by in-orbit ionizing radiation 
even after spikes induced by ionizing particles are removed.
The effects are described as  changes mostly in the offset of detector output, 
but not in the gain.
We conclude that the changes in the offset are caused mainly by increase in dark current. 
We establish a method to correct these ionizing radiation effects. 
The method is essential to improve the quality and to increase the sky coverage of the {\it AKARI} mid-infrared all-sky-survey map.
\end{abstract}
%\keywords{\textcolor{blue}{Si:As infrared detector, ionizing radiation, AKARI,  all-sky survey}}

\section{Introduction}

The effects of high-energy ionizing particles (hereafter we simply call them ionizing radiation effects) on infrared detectors of extrinsic semiconductors are serious for infrared astronomical observations in space.
The ionizing radiation effects cause changes in detector sensitivity and after-effects 
which can last for a certain amount of time after the particle hit.
For infrared astronomical satellites such as {\it Infrared Space Observatory} \citep[ISO;][]{Kessler_1996} and 
{\it Infrared Astoronomical Satellite} \citep[IRAS;][]{Neugebauer_1984}, it was difficult to perfectly remove the ionizing radiation effects from their  observational data \citep[e.g.][]{IRAS_radiation,ISO_Nieminen,ISO_Heras}.
This problem is particularly serious in the south Atlantic anomaly (SAA).
%is the region that radiation belt, composed of high-energy electrons and protons trapped by the magnetic field surrounding the earth, falled low altitude in the sky over Brazil.
The SAA has the local depression in the Earth magnetic field intensity over Brazil, where many high-energy protons and electrons are trapped even at low sky altitudes.
When a satellite passes through the SAA, a great amount of ionizing radiation is irradiated onto detectors.

{\it AKARI}, the first Japanese infrared astronomical satellite, was launched on February 21, 2006 (UT) 
into a sun-synchronous polar orbit at an altitude of 700 km with a period of about 100 minutes per revolution \citep{Murakami}.
{\it AKARI} made an all-sky survey in the six photometric bands centered at the wavelengths of 9 and 18 $\micron$ with the Infrared Camera (IRC; Wada et al. 2003; Onaka et al. 2007) and 65, 90, 140, and 160 $\micron$ with the Far-Infrared Surveyor (FIS; Kawada et al. 2007) within a period of almost one and a half years \citep{Kataza_2010,Yamamura_2010,MirCat}.
The all-sky survey was performed
during the lifetime of liquid helium cryogen 
between May 8, 2006 and August 28, 2007.
As a result, {\it AKARI} surveyed more than
96 and 97 \% of the whole sky twice or more in the S9W and L18W bands, respectively.
During the survey, {\it AKARI} passed through the SAA three or four times a day.

For the far-infrared survey, the {\it AKARI}/FIS adopted Ge:Ga photoconductive devices, which made the all-sky surveys in the 65, 90, 140, and 160 $\micron$  bands.
%In general, extrinsic photoconductors are very sensitive to particle hitting;
The Ge:Ga detectors were seriously influenced by the ionizing radiation, which 
caused undesirable changes in detector responsivity and degradation in noise performance \citep{FIS,Radiation}.
Thus a curing operation was performed every time the satellite passed through the SAA.
The FIS did not carry out pointed observations during the passage of the SAA.

The mid-infrared channels of the {\it AKARI}/IRC, which made the all-sky surveys at the 9 $\micron$ and 18 $\micron$
bands, adopted the Si:As/CRC-744 infrared arrays manufactured by Raytheon \citep{Lum_1993,Estrada_1998,Ando_2003}.
The detector is of an impurity band conduction (IBC) type \citep{Szmulowicz_1987}, which is known to be little influenced by ionizing radiation unlike the above non-IBC-type Ge:Ga detectors \citep{Gordon_2005}.
%\citep[e.g.][]{}
However, spikes are inevitably induced by ionizing particle hitting events in the detector output, which resemble the detection of astronomical point sources.
Therefore, algorithm was developed to remove the spikes induced by ionizing particles \citep{ScanOpe,MirCat}.
So far, it has not been quantitatively investigated whether or not there are any significant changes in 
detector performance other than the spikes due to in-orbit ionizing radiation.

In this paper, we evaluate ionizing radiation effects on the Si:As IBC mid-infrared detector on board {\it AKARI}.
We find that the detector is significantly affected by in-orbit ionizing radiation 
even after spikes induced by ionizing particles are removed.
The effects are described as changes mostly in the offset of detector output, but not in the gain, where 
the offset value increases with the particle hitting rate.
We establish a method of correction, by which we can utilize the data affected by the ionizing radiation  
to make diffuse sky maps.

%; previously such data were simply discarded.

\section{Data reduction}

\begin{figure*}[t]
\epsscale{.9}
\includegraphics[width=17cm]{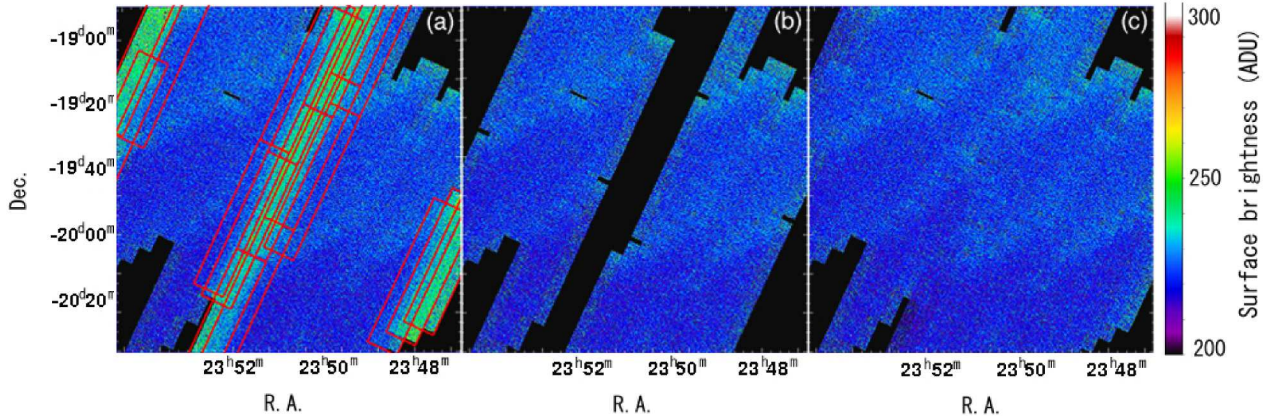}
\caption{{\it AKARI} 18\,$\mu$m maps 
of the $1^{\circ}.7\times1^{\circ}$.7 area 
centered at (R.A., Dec.)=(357$^\circ$.7, $-$19$^\circ$.7):
%The bin size is $6\farcs24\times6\farcs24$.
(a) original image created from data including those affected by the ionizing radiation inside the SAA,
(b) the image created by not using the SAA-affected data, and
(c) the image created after correcting the SAA-affected data by using the correction method presented in this paper.
The red boxs in the panel (a) show the image units (see text) with particle hitting rates higher than 5 events unit$^{-1}$, which are defined as SAA-affected data.
%(a)the image created by using the data taken while the satellite passed the SAA, 
%and 
%(b)the image after removing such data, and 
%(c)the image after correcting such data (reference text body Sec.\ref{sec:corr}).
}
\label{fig:1stmap}
\end{figure*}

We use the all-sky survey data in the 9 $\micron$ and 18 $\micron$ bands to investigate ionizing radiation effects 
on the Si:As IBC detectors.
The contents of this paper are based on the 18 $\micron$ data; for the 9 $\micron$ data, similar results are obtained and the same technique as described below can be applied.

In the all-sky survey observations,
two out of 256 rows in the $256\times256$ detector array format are operated 
in a continuous non-destructive readout mode \citep{ScanOpe}.
The pixel scales are $2\farcs34\times2\farcs34$ for the 9 $\micron$ band and $2\farcs51\times2\farcs39$ for the 18 $\micron$ band. 
The sampling rate is set to be 22.72 Hz (one sampling per $4.4$$\times$$10^{-2}$ s), 
while the scan speed of the satellite in the survey observation mode is $216''$ ${\rm s}^{-1}$.
The effective pixel size for a single scan by a single row is
$9\farcs36\times9\farcs36$ for the 9 $\micron$ band 
and $10\farcs4\times9\farcs36$ for the 18 $\micron$ band.
The in-scan pixel size is fixed by the sampling rate, while the outputs of four neighboring pixels are coadded to meet 
the down-link rate requirements and still make effective observations.
Finer spatial resolutions comparable to the point spread functions (PSFs) of $5\farcs5$ and $5\farcs7$ in FWHM for the 
9 $\micron$ and the 18 $\micron$ band, respectively \citep{IRC}, 
are obtained in the data processing by combining images produced by the two rows \citep{MirCat}.

All the pixels were reset at a rate of 0.074 Hz
(one reset per 306 samplings) to discharge the photo-current integrated in detector capacitance.
The region of $10' (=2\farcs34\times256$ pixels) $\times48'$(=$216''$/0.074 Hz) was surveyed in this reset interval.
Here and hereafter we define this region as an image unit.
After we process the data with the pipeline used in generating the point source catalog,
we calculate averaged pixel intensities per image unit.
The spikes induced by ionizing particle hitting are detected by millisecond confirmation 
by using the redundancy of the 2 rows, and removed from the data \citep{MirCat}.
We define the number of the spikes detected per image unit as particle hitting rate per image unit.

Figure~\ref{fig:1stmap} presents an example of {\it AKARI} 18 $\micron$ maps 
of the $1^{\circ}.7\times1^{\circ}.7$ area centered at (R.A., Dec.)=(357$^\circ$.7, $-$19$^\circ$.7), 
where we have already removed the spikes induced by ionizing particles from the data.
% described above.
%by using two independent data obtained by the two rows \citep{MirCat}.
The size of the image bin is set to be $6\farcs24\times6\farcs24$.
Figure~\ref{fig:1stmap}a is the image created by using the data including those taken during the passage of the SAA, 
while Fig.~\ref{fig:1stmap}b is the image created by not using such data.
%The panel (a) is shown bright streaks and the panel (b) is not found.
The bright stripes are seen in the original image (Fig.~\ref{fig:1stmap}a).
However, after the data taken during the passage of the SAA are excluded, the stripes disappear from the image (Fig.~\ref{fig:1stmap}b).
%This data were observed on passage through the SAA and 
%we ,thus, find that the ionizing radiation has significant effects on the IBC detector more than producing discrete spikes.
We thus find that the ionizing radiation
has significant effects on the IBC detector output
even after the spikes are removed.
Here and hereafter, the SAA-affected data are defined as those included in image units 
with particle hitting rates higher than 5 events unit$^{-1}$. 

\begin{figure}[h]
\epsscale{.9}
%\plotone{delta_z_1.eps}
%\includegraphics[width=17cm]{incomplete_03.eps}
\includegraphics[width=17cm]{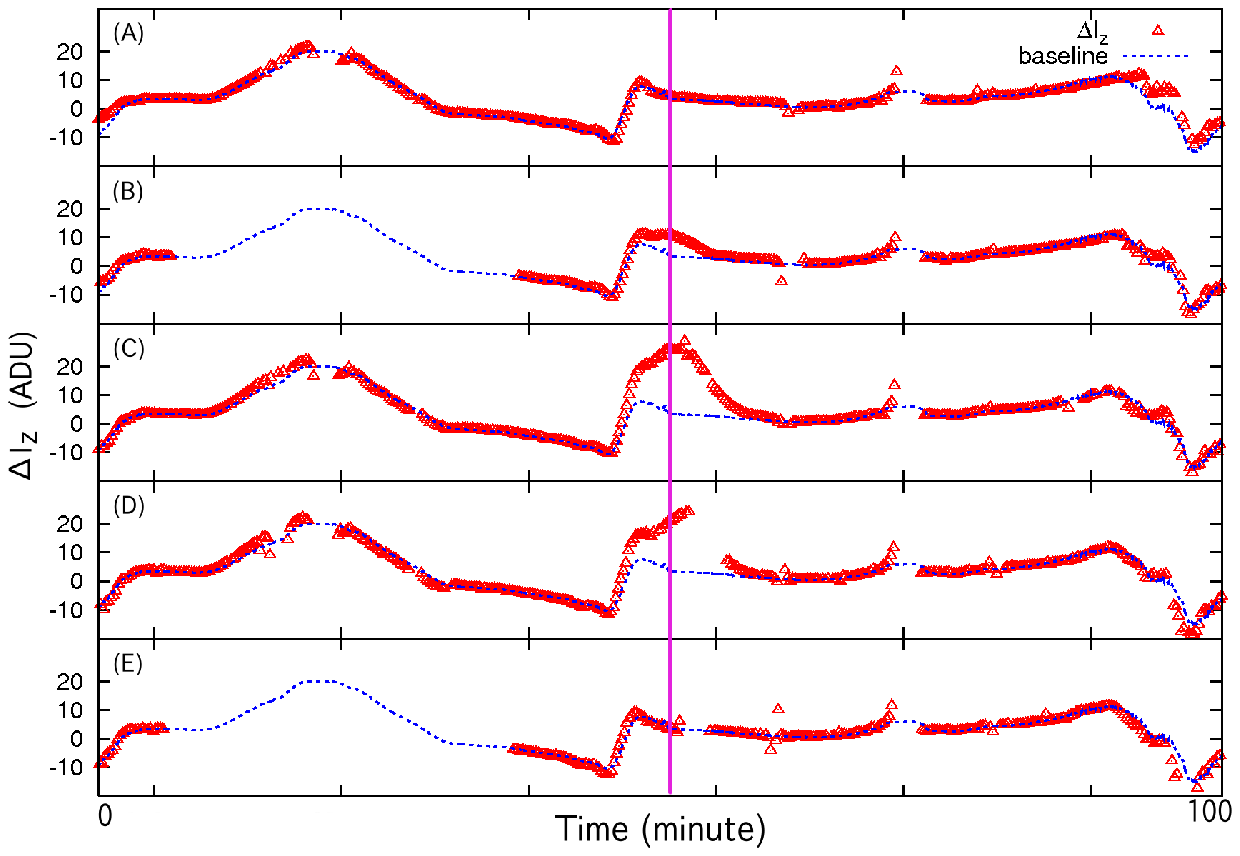}
\caption{
Difference of the foreground emission between two seasons, ${\Delta}$$I_z$, plotted as a function of time during one revolution of the {\it AKARI} satellite for each of 5 sequential orbits.
%We use two datasets taken on June 10--11, 2006 and December 10--11, 2006 (UT).
The triangles show observed ${\Delta}$$I_z$, part of which are affected by the SAA or lost by pointed observations.
The dashed lines show ${\Delta}$$I_z$ where the SAA-affected or lost data are estimated through interpolation by the observational ${\Delta}$$I_z$ for 24 hours.
The perpendicular line roughly indicates the center of the SAA region.
%The panels A and E correspond to orbits before and after the satellite passes the SAA, respectively, 
%while the panels B, C, and D correspond to orbits when the satellite passes the SAA.
See text for details.
}
\label{fig:delta_z}
\end{figure}

As shown in Fig.~\ref{fig:1stmap}a, the ionizing radiation apparently increases the output level of the detector, 
which can be explained by change in either offset or gain of the output.
%In the next section,
%we investigate the cause of the increase in the output level.
%Fortunately,
Because {\it AKARI} observed the same region twice or more 
and the IRC was continuously operated around the SAA, 
%we can compare the data taken inside the SAA with the data outside the SAA for the same sky region.
we can compare SAA-affected and non-affected intensities to quantitatively evaluate the change by the ionizing radiation 
by using two datasets taken at the same sky position in different seasons.

However, we cannot simply compare the data in different seasons
because the Zodiacal light emission,
a dominant foreground emission component in the mid-IR wavelength region, is known to 
show seasonal variations due to asymmetry in distribution of Zodiacal dust with respect to the ecliptic plane and the circum-solar dust ring \citep{Kelsall}.
Thus, we first develop a method to evaluate the seasonal variation of the foreground emission.

The difference of the foreground emission ($\Delta I_{\rm z}$) 
between two seasons due to the variation of the Zodiacal light 
is represented as,

\begin{equation}
\Delta I_{\rm z} = I_{\rm 1} -  I_{\rm 2} ,
\end{equation} 
where $I_{\rm 1}$ and $I_{\rm 2}$ are the averaged pixel intensities of image units at the same sky position 
observed in the seasons 1 and 2, respectively.
Both $I_{\rm 1}$ and $I_{\rm 2}$ are not affected by the SAA.
Figure~\ref{fig:delta_z} shows  an example of ${\Delta}$$I_z$ as a function of the orbital phase of {\it AKARI} 
(the orbital period of {\it AKARI} is approximately 100 minutes).
The time corresponds to the orbital phase, with which 
the terrestrial latitude of the satellite position and the ecliptic latitude of the observed sky change.
The $I_{\rm 1}$ values are taken from data obtained on June 10, 2006, while 
the $I_{\rm 2}$ values are taken from data obtained half a year later, on December 10, 2006.
Because the orbit of {\it AKARI} is sun-synchronous, almost the same sky positions are covered in these two observations.
The plots A-E in Fig.~\ref{fig:delta_z} correspond to data in 5 sequential orbits.
The maximum distance of sky positions at the same orbital phase between two adjacent orbits is 4 arcmin, and 
that between orbits within 24 hours is one degree.
The plots B, C, and D correspond to data in the orbits passing the SAA.
Large discrepancies between the estimated and the observed $\Delta I_{\rm z}$ are seen during the passage of the SAA, 
where the data cannot be used in estimating $\Delta I_{\rm z}$.
They are due to the ionizing radiation effects.
By assuming that $\Delta I_{\rm z}$ has a smooth spatial distribution within one degree and a small time variation within 24 hours,
non-SAA-affected levels of $\Delta I_{\rm z}$ are 
estimated by interpolating from the neighboring orbit data along the cross-scan direction. 
%and the data observed before and after the satellite passes through the SAA along the scan direction.
We do not use data 
when the resulting $\Delta I_{\rm z}$ shows a time variation faster than 0.44 ADU per minute.
By using these plots, we can distinguish the effects of ionizing radiation from the seasonal variation of the foreground emission. 
%Based on the spike rates, we searched SAA-affected sky positions for the former dataset, 
%where the latter data are not affected.

%We corrected the baseline levels of the latter dataset for the zodiacal light to match the former ones.
%Based on the spike rates, we searched SAA-affected sky positions for the former dataset, 

%where the latter data are not affected.
%Then, based on the spike rates, we searched SAA-affected sky positions for the former dataset, 
%where the latter data are not affected.
%This sample contains various sky brightness 
%and various particle hitting rates.

\begin{figure}[h]
\epsscale{.9}
\includegraphics[width=15cm]{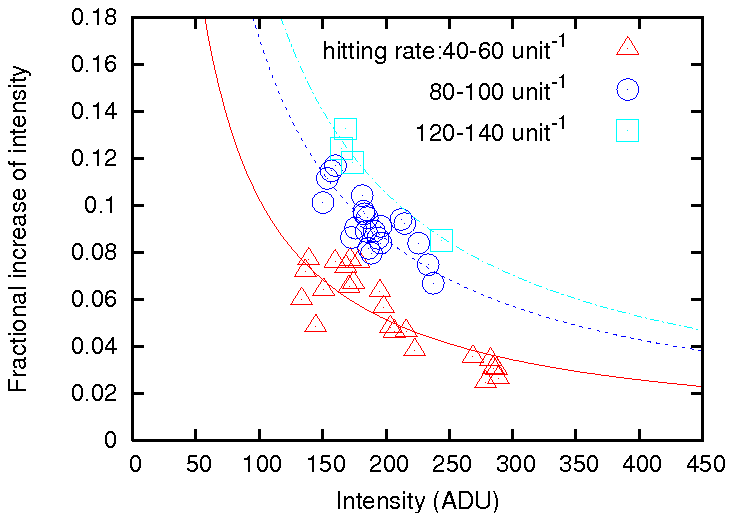}
\caption{Fractional increase of the output level 
due to ionizing radiation effects 
($f(I^{SAA}_{\rm 1})$ in eq.2) 
plotted
against the averaged pixel intensities 
inside the SAA ($I^{SAA}_{\rm 1}$).
We use the same datasets as those used in Fig.~\ref{fig:delta_z}.
The data points with the different ranges of particle hitting rates are 
indicated by different marks: the triangles for 40--60 {unit}$^{-1}$, circles for 80--100 {unit}$^{-1}$, 
squares for 120--140 {unit}$^{-1}$, and fitted by different hyperbolic curves.
}
\label{fig:off-gain}
\end{figure}

\section{Results}
\subsection{Behavior of ionizing radiation effects}

As shown in the previous section, the output levels of the detector increase systematically in the SAA
even after the spikes are removed.
Therefore we examine how they increase, i.e. by change in offset or gain of the output.

The fractional increase of the output level due to the SAA, $f(I^{SAA}_{\rm 1})$, 
is derived from the equation,
\begin{equation}
  f(I^{SAA}_{\rm 1})
  = \frac{(I^{SAA}_{\rm 1}-I_{\rm 2}-\Delta I_{\rm z})}{I^{SAA}_{\rm 1}} ,
\end{equation}
where $I^{SAA}_{\rm 1}$ is the averaged pixel intensity of an image unit observed in the season 1 during the passage of the SAA.
If the ionizing radiation effect is described by the change in the gain (1+$\alpha$), 
$f(I^{SAA}_{\rm 1})$ is reduced to,

\begin{equation}
f(I^{SAA}_{\rm 1})
= 
\frac
{(1+\alpha)I_{\rm 1} - I_{\rm 2}-\Delta I_{\rm z}}
{(1+\alpha)I_{\rm 1}}
 = 
\frac
{\alpha}
{1+\alpha} .
\end{equation}
The scatter plot of  $f(I^{SAA}_{\rm 1})$ versus $I^{SAA}_{\rm 1}$ should follow
the line parallel to a horizontal axis, if $\alpha$ is constant.
On the other hand,
if the ionizing radiation effect is described by the offset ($\beta$),
the scatter plot follows a hyperbolic function as,

\begin{equation}
f(I^{SAA}_{\rm 1})
= 
\frac
{(I_{\rm 1} + \beta) - I_{\rm 2} - \Delta I_{\rm z} }
{(I_{\rm 1}+\beta)}
=
\frac
{\beta}
{I^{SAA}_{\rm 1}} .
\end{equation}
Therefore, 
we can judge whether the ionizing radiation effect is the gain or the offset
from the behavior of $f(I_{\rm 1}^{\rm SAA})$ with respect to $I_{\rm 1}^{\rm SAA}$.

\begin{figure*}[t]
\epsscale{.9}
\includegraphics[width=15cm]{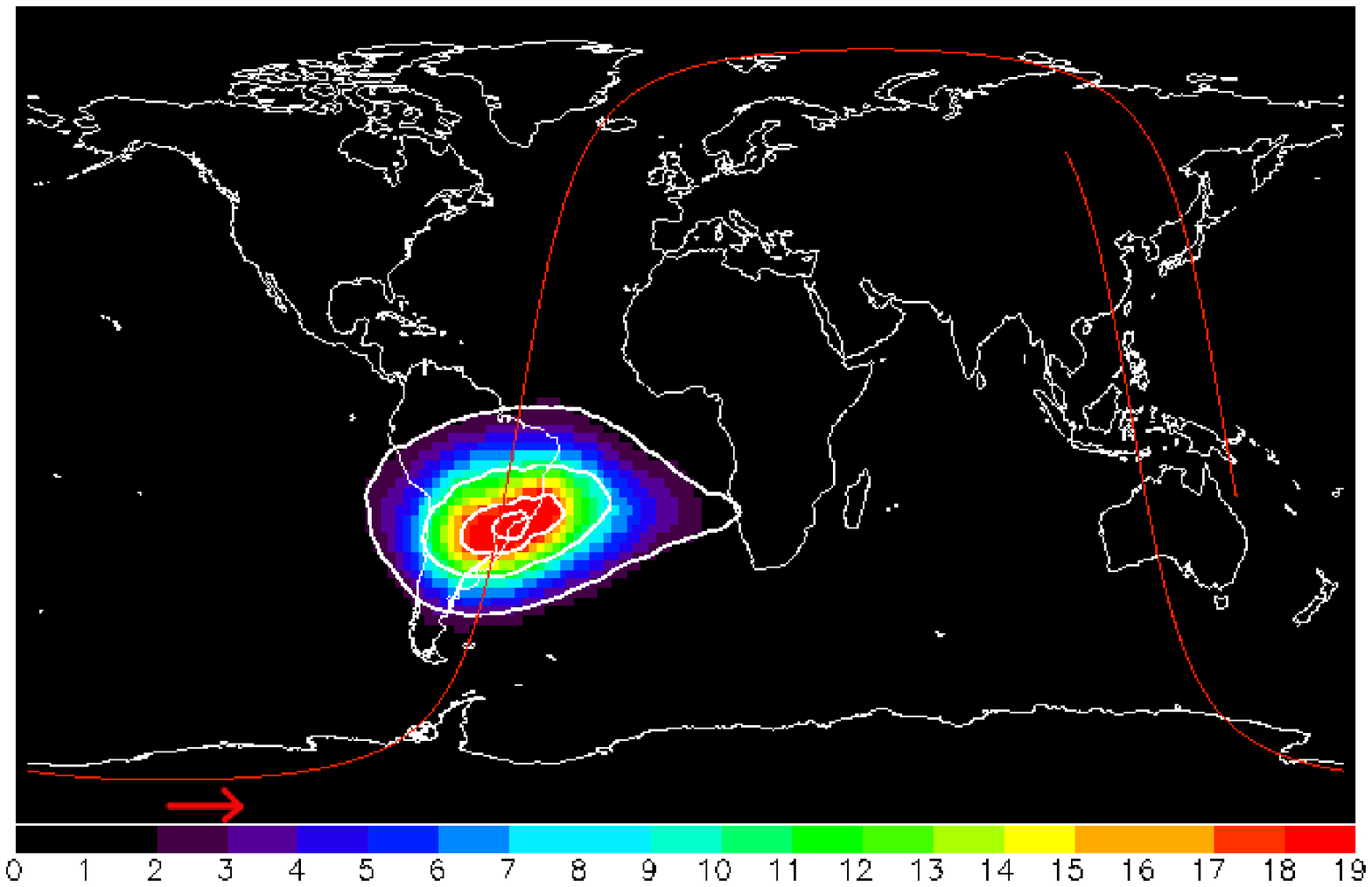}%Fig.~\ref{fig:1stmap}
\caption{Distribution map of the offset value 
of the detector output ($I^{SAA}-I$) shown in the terrestrial coordinates.
The contours show the distribution of the particle hitting rate, which are drawn by the 4 levels from 5 to 120 unit$^{-1}$ on a linear scale.
%The SAA is located around (lon, lat) = ($-$45$^\circ$, $-$30$^\circ$). 
The lowest contour corresponds to the boundary of the SAA defined in this paper.
The red line shows the trajectory of {\it AKARI} that passes the SAA 
in the direction as indicated by the arrow.
}
\label{fig:earth}
\end{figure*}

Figure~\ref{fig:off-gain} shows the plot
of $f(I_{\rm 1}^{\rm SAA})$ versus $I_{\rm 1}^{\rm SAA}$ 
for the three different ranges of particle hitting rates. 
Because these plots are better fitted 
by hyperbolic curves than by lines parallel to the horizontal axis, 
we conclude that the ionizing radiation effect is explained mostly by the change in the offset of detector output, 
but not in the gain.

Figure~\ref{fig:earth} compares the distribution map of the offset value of the detector output ($\beta$=$I^{SAA}$$-$$I$) 
with the contours of the particle hitting rate in the terrestrial coordinates.
%\textcolor{red}{However, as indicated in Fig.2,
%the offset value of the averaged pixel intensities ($\beta$=$I^{SAA}$$-$$I$) 
%exhibits relatively larger error when the baseline level changes rapidly.
%We excluded such data (the slope of the baseline exceeds 0.1 or underruns -0.1 (ADU/unit)).
%%But the offset value of the detector output ($\beta$=$I^{SAA}$$-$$I$) derived from the data, which observed in the time that %the time derivation of non-SAA-affected levels of $\Delta I_{\rm z}$ 
%%are greater than absolute value of 0.1, are removed.
%%しかし、惑星間塵（ADU）の時間変化量が絶対値0.1以上の点から得られたβ＝I＿SAA　ー　I　は、誤差が大きくなるため、取り除いてある。
%}
%
The distribution map of the offset is spatially correlated well with the contour map of the particle hitting rate, 
and does not elongate along the satellite scan direction, 
indicating that after-effects due to the ionizing radiation are negligible, at least, 
on time scales longer than a few minutes.

\begin{figure}[h]
\epsscale{.9}
%\plotone{saa_hit_VS_uwanose.eps}
\plotone{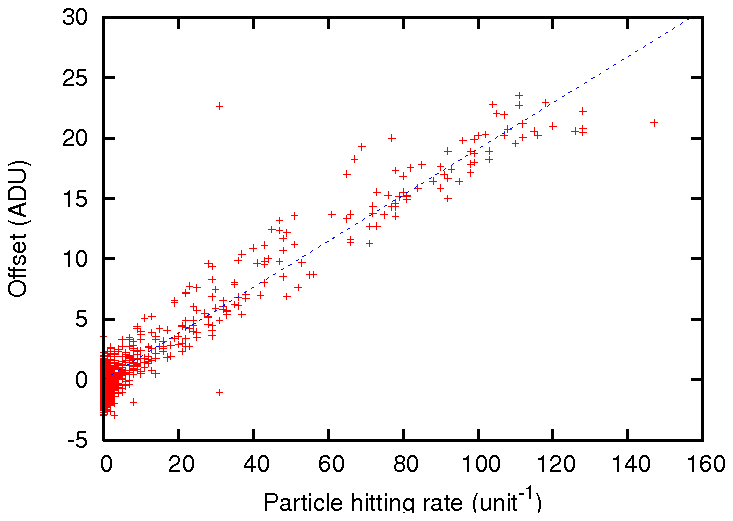}
\caption{Offset values of the averaged pixel intensities ($I^{SAA}-I$) plotted as a function of the particle hitting rate.
We use the same datasets as those used in Fig.~\ref{fig:delta_z}.
The dashed line shows a linear relation between the offset values and the particle hitting rates obtained by a line fittng.
%The offset is the average of difference between SAA-affected and non-affected intensities 
%for all the pixels of an image unit. %with a size of $10'\times45'$ and a pixel scale of $6\farcs24\times6\farcs24$.
%in case that the pixel scale is $6\farcs24\times6\farcs24$.
%The count rate indicates
%the number of the significantly detected spikes by the detector per 13.5\,sec
%(during the time covering $10'\times45'$ area).
%The dataset shown in this plot was taken on 10 June 2006.
}
\label{fig:vs_hit}
\end{figure}

\subsection{Correction of ionizing radiation effects}

We evaluate the ionizing radiation effects 
quantitatively to formulate a correction method.
Figure~\ref{fig:vs_hit} plots offset values ($\beta$=$I^{SAA}$$-$$I$)
as a function of the particle hitting rate.
%for the same dataset as used to create the plot in Fig.~\ref{fig:delta_z}.
The plot shows a linear relation between the offset values and particle hitting rates.
%We interpeted this relation by the linear function,
%\begin{equation}
%  {\rm offset} = C \times {\rm count\_rate}
%\end{equation}
We obtain the linear coefficient of $0.191\pm0.001$
by a line fitting.

\begin{figure*}[t]
\epsscale{.9}
\includegraphics[width=17cm]{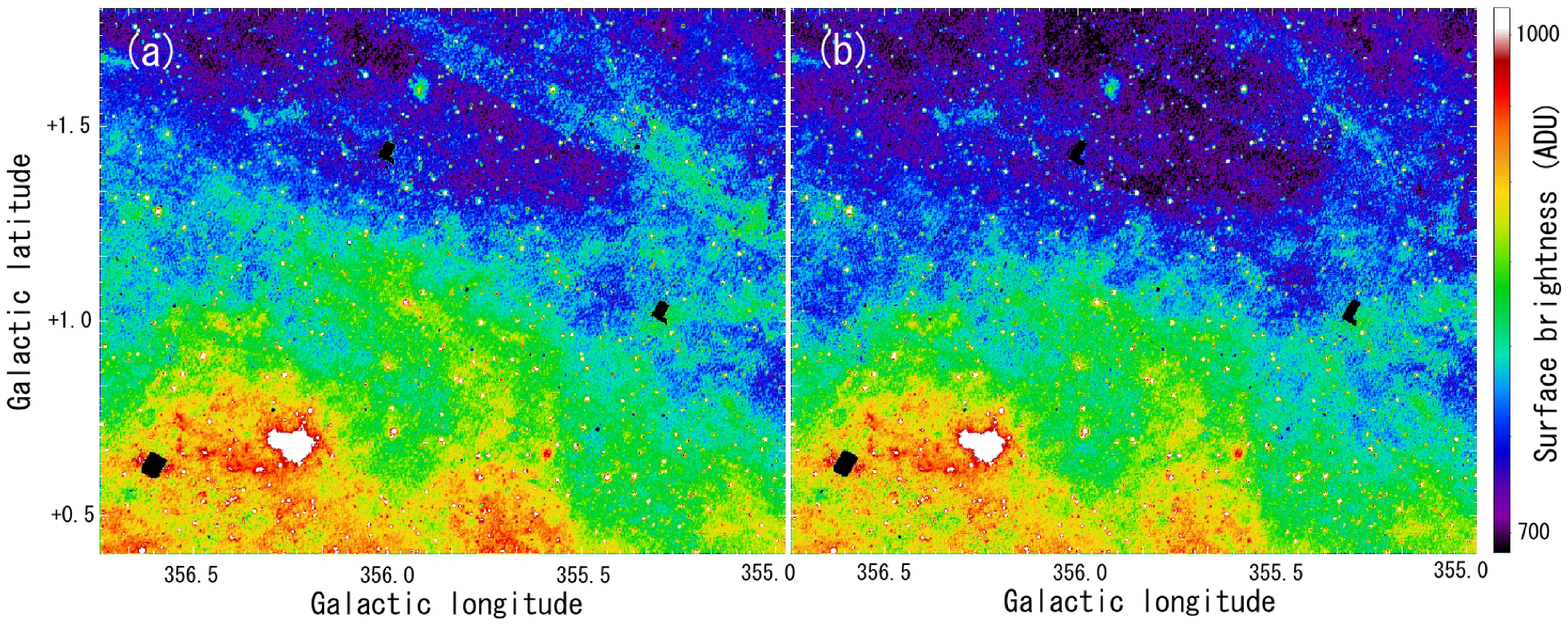}
\caption{{\it AKARI} 18\,$\mu$m maps 
%of the $1^{\circ}.7\times1^{\circ}.4$ area 
near the Galactic center:
%(l, b)=(355$^\circ$.9, 1$^\circ$.1):
(a) original image created by including the data taken inside the SAA,
(b) the image created after correcting the SAA-affected data by using the correction method presented in this paper.
}
\label{fig:galactic}
\end{figure*}

Thus in order to make diffuse maps, we first obtain a particle hitting rate every image unit to be used.
Then as a function of the particle hitting rate, 
we derive the corresponding offset values from the line in Fig.~\ref{fig:vs_hit}, 
and subtracted them from the image units affected by the SAA.
As a result, we derive such a clean image as shown in Fig.~\ref{fig:1stmap}c, 
which demonstrates the effectiveness of this correction method.
We also apply this method to another region near the Galactic center (Fig.~\ref{fig:galactic}).
As can be confirmed from the figure, this method works out well in making diffuse maps with brighter intensities and larger dynamic ranges of brightness.
Thus by making the most use of the data affected by the ionizing radiation with this method, 
we can increase the sky coverage and the quality of the {\it AKARI} mid-IR 
all-sky-survey map, especially the Galactic plane map.

\section{Discussion}

\begin{figure}
\epsscale{.9}
%\includegraphics{SAA_hikaku.eps}
%\plotone{F0300365516_4NL_hist.eps}
%\plotone{Intensity_hist.eps}
%\rotatebox{-90}{\includegraphics{SAA_hikaku.eps}}
\rotatebox{-90}{\includegraphics{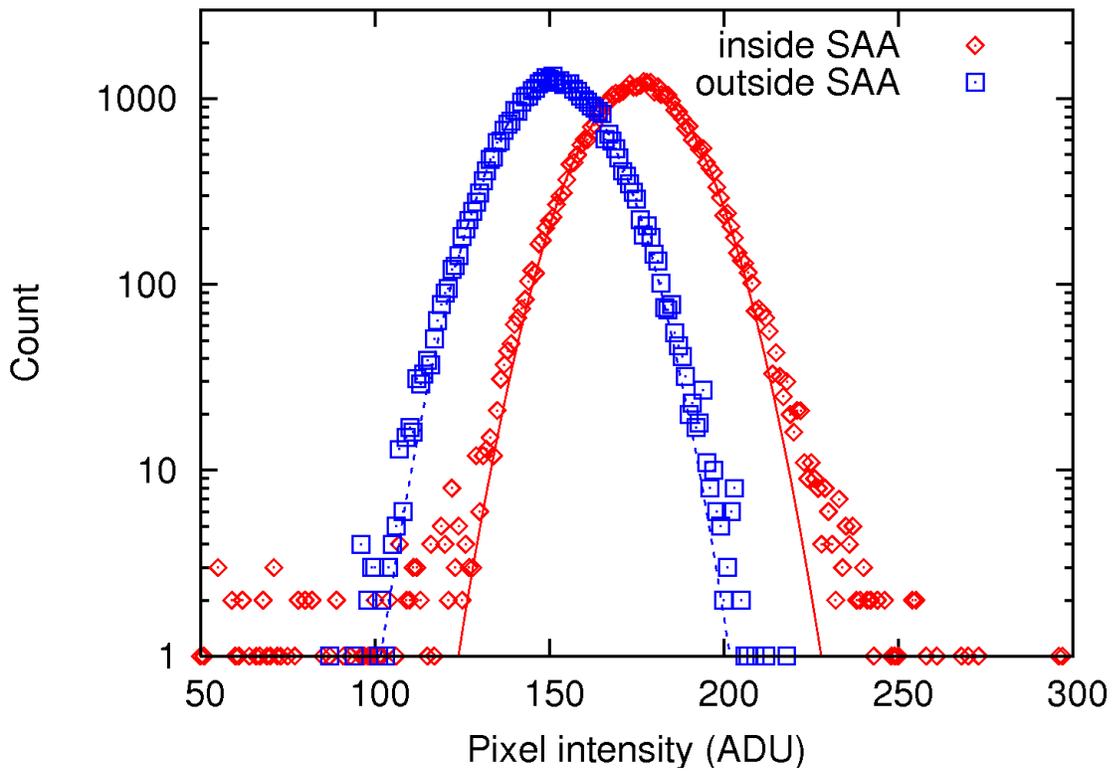}}
\caption{Histograms of pixel intensities 
in the two image units viewing the same blank sky of 
%$1^\circ\times1^\circ$ area
$10'\times48'$ area
around (R.A., Dec.) = (159$^\circ$.8, $-$25$^\circ$.0).
The diamonds indicate the data taken inside the SAA,
while the squares indicate the data taken outside the SAA.
}
\label{fig:gauss2}
\end{figure}

\begin{table}
\caption{Result of Gaussian fitting to the histograms in Fig.~\ref{fig:gauss2}.}
\label{table:sigma}
\center
\begin{tabular}{lcc}
\hline
& Center (ADU)& Width (ADU) \\
\hline \hline
Outside SAA & 151.70$\pm$0.04 & 13.26$\pm$0.04 \\
Inside SAA     & 175.90$\pm$0.05 & 13.82$\pm$0.05 \\
\hline
\end{tabular}
\end{table}

\begin{figure}
\epsscale{.9}
%\rotatebox{-90}{\includegraphics{surface_sigma.eps}}
\rotatebox{-90}{\includegraphics{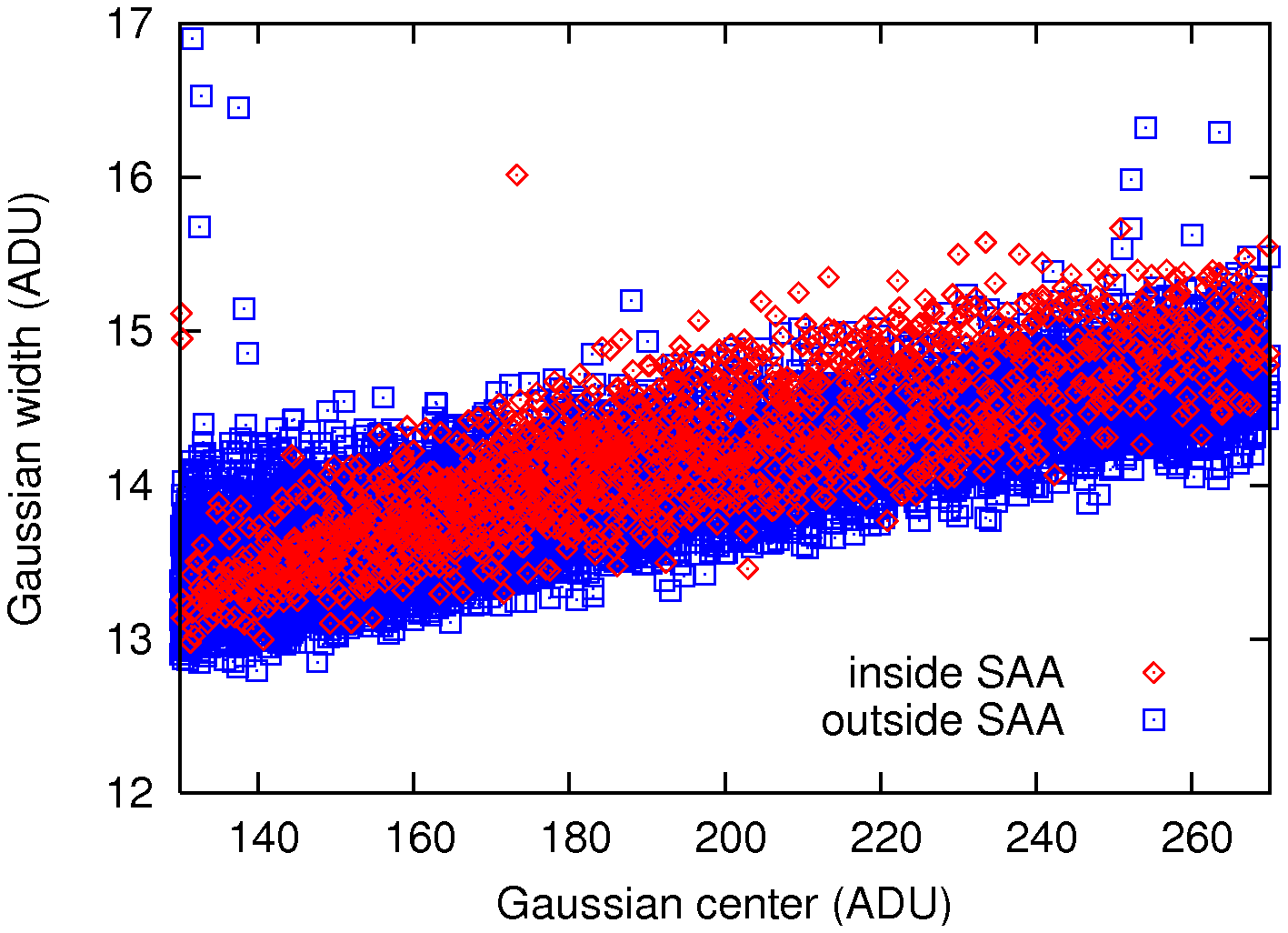}}
%\includegraphics[width=17cm]{surface_sigma.eps}
%\plotone{F0300365516_4NL_hist.eps}
%\plotone{Intensity_hist.eps}
\caption{
Gaussian widths plotted against the Gaussian centers, both obtained by fitting to the histgrams of pixel intensities of image units.
One data point is obtained from one histgram such as shown in Fig.~\ref{fig:gauss2}.
%Gaussian width of the histgram of the surface brightness 
%of the images plotted as a function of the surface brightness.
The data are taken between June 08 and June 14 2006 (UT).
The diamonds indicate the data taken inside the SAA,
while the squares indicate the data taken outside the SAA.
}
\label{fig:sigma}
\end{figure}

What causes the offset of the output level during the passage of the SAA?
Possible scenarios are:
(1) an increase in the number of very small spikes below the detection level,
(2) an increase in dark current, or
(3) a drift of the offset level of readout electronics.
The last one is due to the change of the gate-source voltage
of a source-follower in the readout integrated circuit (ROIC) hybridized to the detector \citep{Wu_1997}, 
which is easily caused by the drift of an array temperature \citep{Ishihara_2003}.
To distinguish them,
we compare 
%not only the average of the surface brightness,
%but also 
the distribution of the pixel intensities 
of the image unit taken inside the SAA with that outside the SAA, 
as shown in Fig.~\ref{fig:gauss2}.

Each distribution is well explained by a Gaussian distribution.
The fitting parameters of the Gaussian center and width are summarized in Table~\ref{table:sigma}, together with their errors.
The ionizing radiation effects make the distribution shift toward higher intensities and increase its width (i.e. noise) significantly.
But it keeps the symmetric shape with respect to the peak,
which indicates that the rise of the averaged intensities inside the SAA
is caused by a slight increase in intensities of all the pixels, 
but not an increase in the number of positive spiky pixels.
We have confirmed the presence of residual small positive and negative spikes 
below the detection limit of the de-spiking algorithm \citep{ScanOpe,MirCat}, 
producing the low-level broad wings in Fig.7.
However such spikes are only a few percent in number, 
which cannot make a significant contribution to the observed peak shift of the distribution.

%It keeps the symmetric shape with respect to the peak, and
%the distribution taken inside the SAA looks broader in the wings than that taken outside the SAA, 
%which caused by increase in the number of very small spikes that cannot be removed by algorithm and the negative spikes caused presumably by saturation of a source-follower of unit cell.
%But the number of these spikes is a few percent of all pixels, 
%which does not affect multitude. 
%Therefore, we conclude that the rise of the averaged intensities inside the SAA
%is caused by a slight increase in intensities of all the pixels, 
%but not an increase in the number of spiky pixels.}

%Furthermore, 
%$\sigma$ of the histogram for the data taken inside the SAA 
%is significantly larger than that taken outside the SAA.
%
%It indicates and explained quantitatively that
%the offset of the surface brightness inside the SAA is caused
%by the actual increase in dark electric current, 
%but not by the simple drift of the offset of the ROIC.

In Fig.~\ref{fig:sigma}, we plot the Gaussian widths against the central intensities, which are obtained by fitting 
a Gaussian profile to each histogram;
by using the data taken between June 08 and June 14 2006 (UT), 
we obtain many histgrams such as shown in Fig.~\ref{fig:gauss2}.
If the offset of the output level is caused by a simple drift of the ROIC, 
the data taken inside the SAA should shift towards higher intensities along the horizontal axis 
while the Gaussian width does not change.
However, the data taken inside the SAA show increases in the Gaussian widths as well as the intensities.
Thus the noise does change due to the ionizing radiation effect in the SAA.
Moreover the overall distributions of the data taken inside the SAA and outside the SAA show 
almost no systematic difference.
Since dark current and photo-current are expected to have similar dependence of noise on intensities, 
Fig.~\ref{fig:sigma} indicates that the offset of an averaged pixel intensity inside the SAA is caused
mainly by the actual increase in dark current, 
but not by a simple drift of the offset of the ROIC.

Therefore, we conclude that the offset of the output level during the passage of the SAA 
is caused mainly by increases in dark current of the detector.

\section{Summary}

We have investigated ionizing radiation effects on the Si:As IBC mid-infrared detector on board the {\it AKARI} satellite.
By using the {\it AKARI} all-sky-survey data, we have found that the detector is significantly affected by in-orbit 
ionizing radiation during the passage of the SAA.
The effects are described as changes mostly in the offset of detector output, but not in the gain.
We have confirmed that the distribution of the offset value shows an excellent spatial correlation with the distribution 
of the particle hitting rate in the terrestrial coordinate map, exhibiting no significant after-effects due to 
the passage of the SAA.
The offset values show a linear correlation with the particle hitting rates.
By utilizing the linear correlation, we have established a method to correct the ionizing radiation effects 
and successfully applied it to the all-sky-survey data, 
resulting in producing better diffuse emission maps.
By investigating the histograms of pixel intensities in image units taken inside and outside the SAA, 
we conclude that the main physical cause of the offset variation is the increase of dark current.
%Si:As IBC detectors are adopted in on-going and is scheduled to be used for future missions \citep[e.g.][]{Mainzer_2005,Love_2005}.
%This research is expected to be useful for future missions. 

\acknowledgments

This research is based on observations with {\it {\it AKARI}},
a JAXA project with the participation of ESA.
This work was supported by
the the Nagoya University Global COE Program, ``Quest for Fundamental Principles in the Universe (QFPU)''
from the JSPS and MEXT of Japan.
We would express many thanks to the anonymous referee 
for giving us useful comments.

{\it Facilities:} \facility{{\it AKARI} (IRC)}

\end{document}